\documentclass[letterpaper,aps,prd,preprint,showpacs,nofootinbib,superscriptaddress]{revtex4-1}
\usepackage{amsmath,amssymb,graphicx}
\usepackage[stable]{footmisc}

\pdfoutput=1

\newcommand{\nc}{\newcommand}
\nc{\postscript}[2] 
{\setlength{\epsfxsize}{#2\hsize}\centerline{\epsfbox{#1}}}
\nc{\non}{\nonumber}
\nc{\hc}{\hbox {h.c.}} \nc{\re}{\hbox {Re}} 
\nc{\mev}{\hbox {MeV}} \nc{\gev}{\;\hbox {GeV}} \nc{\tev}{\;\hbox {TeV}}
\def\lsim{\mathrel{\raise.3ex\hbox{$<$\kern-.75em\lower1ex\hbox{$\sim$}}}}
\def\gsim{\mathrel{\raise.3ex\hbox{$>$\kern-.75em\lower1ex\hbox{$\sim$}}}}

\nc{\etal}{{\it et al.}}
\nc{\Lsp}{\;\;\;\;\;\;\;\;\;\;}  \nc{\LLLsp}{\lspace \lspace}
\nc{\lsp}{\;\;\;\;\;\;}
\nc{\spac}{\;\;\;}
\nc{\noi}{\noindent}
\nc{\beq}{\begin{equation}}   \nc{\eeq}{\end{equation}}
\nc{\bea}{\begin{eqnarray}}   \nc{\eea}{\end{eqnarray}}
\nc{\baa}{\begin{array}}      \nc{\eaa}{\end{array}}
\nc{\bit}{\begin{itemize}}    \nc{\eit}{\end{itemize}}
\nc{\ben}{\begin{enumerate}}  \nc{\een}{\end{enumerate}}
\nc{\bce}{\begin{center}}     \nc{\ece}{\end{center}}

\def\calQ{{\cal Q}}
\def\calU{{\cal U}}

\def\sq2{\sqrt{2}}

\def\ph{\varphi}

\def\m4{m^4(\ph)}
\def\mn2{m_n^2}

\def\v5{V^{(5)}}

\def\baa{\begin{array}}
\def\eaa{\end{array}}
\def\l{\left}
\def\r{\right}
\def\d{\partial}

\begin{document}

\begin{flushright}
 \mbox{\normalsize \rm CUMQ/HEP 171}\\
\end{flushright}

\vskip 20pt

\title{
Higgs Bosons in Warped Space, from the Bulk to the Brane}

\author{Mariana Frank\footnote{mariana.frank@concordia.ca}}
\affiliation{Department of Physics, Concordia University\\
7141 Sherbrooke St. West, Montreal, Quebec,\\ CANADA H4B 1R6
}
\author{
Nima Pourtolami\footnote{n\_pour@live.concorida.ca}}
\affiliation{Department of Physics, Concordia University\\
7141 Sherbrooke St. West, Montreal, Quebec,\\ CANADA H4B 1R6
}
\author{Manuel Toharia\footnote{mtoharia@physics.concordia.ca}}
\affiliation{Department of Physics, Concordia University\\
7141 Sherbrooke St. West, Montreal, Quebec,\\ CANADA H4B 1R6
}

\date{\today}

\begin{abstract}

In the context of warped extra-dimensional models with all fields
propagating in the bulk, we address the phenomenology of a bulk scalar
Higgs boson, and calculate its production cross section at the LHC as well
as its tree-level effects on mediating flavor changing neutral currents. 
We perform the calculations based on two different approaches. 
First, 
we compute our
predictions analytically by considering all the 
degrees of freedom emerging from the dimensional reduction (the
infinite tower of Kaluza Klein modes (KK)). In the second approach, we
perform our calculations numerically by considering only the effects
caused by the first few KK modes, present in the 4-dimensional effective theory.
In the case of a Higgs leaking far from the brane, both approaches give
the same predictions as the effects of the heavier KK modes
decouple. However, as the Higgs boson is pushed towards the TeV brane, the two
approaches seem to be equivalent only when one includes heavier and
heavier degrees of freedom (which do not seem to decouple). To
reconcile these results 
it is necessary to introduce a type of higher derivative operator
which essentially encodes the effects of integrating out the heavy KK
modes and dresses the brane Higgs so that it looks just like a bulk
Higgs.

\end{abstract}

\pacs{11.10.Kk, 12.60.Fr, 14.80.Ec}

\maketitle

\section{Introduction}
\label{sec:intro}

Warped extra dimensional models have become very popular because they
are able to address simultaneously two intriguing issues within the
Standard Model (SM): the hierarchy problem and the mass/flavor problem. 
They were originally introduced to treat the first issue \cite{RS} in a
setup where the SM fields were all localized at one boundary of the
extra dimension. Later it was realized that by allowing fields to
propagate into the bulk, different geographical localization of
fields along the extra dimension could help explain the observed
masses and flavor mixing among quarks and leptons \cite{bulkSM, fermionsbulk}. Flavor
bounds and precision electroweak tests put pressure on the
mass scale of new physics in these models \cite{bounds}, but extending
the gauge groups and/or matter content (e.g. \cite{extensions}) or by slightly modifying the spacetime warping of the
metric (e.g \cite{softmodels}), it is possible to keep the 
new physics scale at the TeV level at the reach of the Large Hadron
Collider (LHC).

Electroweak symmetry breaking can still happen via a standard Higgs mechanism
in these scenarios (although it can also be implemented as
as Pseudo-Nambu-Goldstone boson (PNGB) \cite{pgbhiggs} or described
within the effective theory formalism
\cite{Giudice:2007fh,Low:2009di}). As the LHC announced the  
discovery of a light Higgs-like particle of a mass around 125 GeV
\cite{LHCHiggs}, it becomes crucial to have a detailed prediction of
the properties of the physical Higgs particle in these models. The
Higgs boson itself must be located near the TeV boundary of the extra
dimension in order to solve the hierarchy problem, and so typically it
is assumed to be exactly localized on that boundary (brane Higgs
scenario). Nevertheless, it is possible that it leaks out into the
bulk (bulk Higgs scenario), and in doing so indirectly alleviate some
of the bounds plaguing these models \cite{AAZ}. 

The calculation of the production cross section of the brane Higgs in these
scenarios has been addressed before \cite{Lillie:2005pt,Falkowski:2007hz,
  Djouadi:2007fm,Bouchart:2009vq,Cacciapaglia:2009ky,ATZ2,Neubert} but we will pay close attention
to the more recent works of \cite{Bouchart:2009vq,ATZ2,Neubert}. The towers of fermion Kaluza-Klein (KK) 
modes will affect significantly the SM prediction and in
\cite{ATZ2} it was found that the Higgs boson production rate can receive
important corrections, either enhancing or suppressing the Standard
Model prediction. The suppression or enhancement depends on the model
parameters considered, in particular on the phases appearing in the
different Yukawa-type operators present in the 5D action.
Previously, the analysis of \cite{Bouchart:2009vq},
in which only the first few modes were considered, gave no
contribution to the rate from the towers of KK fermions.
Finally, the analysis of \cite{Neubert} seems to indicate that with
just a few KK modes a substantial effect is obtained, but of opposite
sign as the one predicted from summing the infinite tower
\cite{ATZ2}.

In this work we consider the effects of allowing the Higgs boson
to propagate in the bulk, with its profile more or less localized
towards the IR brane depending on the value of the mass parameter
$\beta$, related to the bulk mass of the 5D Higgs field.

To keep matters as simple as possible we will set up a model containing a
single family of up-type 5D fermions along with a bulk Higgs
scalar. Generalization to a more realistic scenario is straight
forward but we prefer to stay as transparent as possible due to the
many subtleties involved in the calculation.

We first compute the contribution of the complete tower of KK fermions
to the Higgs production cross section as well as to the tree-level
shift happening between the light fermion 
mass and its Yukawa coupling (leading to flavor violating
couplings when considering three fermion families). These
calculations, as outlined in \cite{ATZ1,ATZ2}, are 
analytically straightforward and allow us to obtain simple and compact
results. We then repeat the same analysis numerically from the
point of view of an effective theory in which only the first few KK
fermions contribute.
We show that for a bulk Higgs with a thickness of the order of
inverse TeV scale, the results obtained are the same as the 
results obtained by summing the complete KK tower (i.e. heavier modes
decouple). Moreover, these results are consistent with the 
predictions obtained in \cite{ATZ1,ATZ2} for the specific case of a brane
localized Higgs.
The two aproaches outlined seem to give different predictions as the bulk Higgs
is continuously pushed towards the brane. It turns out that in order to
maintain the consistency of both approaches we need to include in the analysis
the effects of a special type of higher order operators. 
After these effects are included, we will come back and address in the
discussion section the differences among the existing calculations in the literature and
stress the importance of including the mentioned higher order operators in the analysis.   

Our paper is organized as follows. In Sec. \ref{sec:setup} we
summarize the simple 5D warped space model used in the calculation. In
Sec. \ref{sec:infinite} we present analytical results for the Higgs
flavor-changing effects (\ref{subsec:infiniteFCNC}) and production
(\ref{subsec:infiniteprod}), using the full tower of KK fermions. We
use numerical methods to calculate the effects of including just a few
KK modes in Sec. \ref{sec:individual}, both for flavor-changing
neutral currents effects   (\ref{subsec:individualFCNC}) and Higgs
boson production (\ref{subsec:individualprod}). We include the effect
of the higher order operator in Sec. \ref{sec:ramanop} and discuss the
misalignment between the Higgs boson profile and its vacuum
expectation value (VEV) in Sec. \ref{sec:vevmisalignment}. We discuss
the significance of our results, compare them to previous analyses
and conclude in Sec. \ref{sec:conclusion}.   We leave some of the
details for the Appendices \ref{apen:I} and \ref{apen:II}.

\section{The Model}
\label{sec:setup}

We consider the simplest 5D warped extension of
the SM, in which we keep the SM local gauge groups and just extend the 
space-time by one warped extra dimension. 

The spacetime metric is the usual Randall-Sundrum form~\cite{RS}:
\bea
ds^2
= \frac{R^2}{z^2}\! \Big(\eta_{\mu\nu} dx^\mu dx^\nu -dz^2\Big),
\label{RS}
\eea
with the UV (IR) branes localized at $z = R$ ($z = R^\prime$).  
We denote the $SU(2)_L$ doublets by ${\cal Q}^i(x,z)$ and the $SU(2)_L$ singlets
by ${\cal U}^j(x,z)$ where $i,j$ are flavor indices and $x$
represents the 4D spacetime coordinates while $z$ represents the extra
dimension coordinate. The fermions are expected to propagate in the
bulk \cite{bulkSM, fermionsbulk}. 

The up-sector fermion action that we consider is therefore
\bea
&&\hspace{-1cm} S_{fermion}=\int d^4x dz \sqrt{g} \left[
{i\over2} \left(\bar{\calQ_i} \Gamma^A {\cal D}_A \calQ_i -
{\cal D}_A \bar{ \calQ_i} \Gamma^A \calQ_i\right) + {c_{q_i}\over R} \bar{ \calQ_i} \calQ_i 
+\right. \non \\
&& \left.{i\over2} \left(\bar{ \calU_j} \Gamma^A {\cal D}_A U_j - 
{\cal D}_A \bar{ \calU_j} \Gamma^A \calU_j\right) + {c_{u_j}\over R}
\bar{ \calU_j} \calU_j \ + \ \left(Y^*_{ij}\ \bar{\calQ_i} {H}  \calU_j + h.c.\right)
\right], \ \ \
\eea with ${\cal D}_A$ being the covariant derivative, and we have
added a Yukawa interaction with a Higgs field $H$ which in principle can be
either brane or bulk localized. From the 5D fermion mass terms one
defines dimensionless parameters $c_{u_i}$, $c_{q_i}$
which are {\it a priori} quantities of ${\cal O}(1)$. The coefficients
$Y^*_{ij}$ have inverse energy units ($1/\sqrt{\Lambda}$) since Yukawa
couplings in 5D are higher dimensional operators.  

After separating 5D fields into left and right chiralities we impose a
mixed ansatz for separation of variables  
\bea
q_L(x,z)   &=&  q_L^0(z) q_L^0 (x)     +\  Q_L^1(z)  \Psi_L^1(x) + ...\, ,\label{sep1}\\
{q}_R(x,z) &=&  q^0_R(z) {u}^0_R(x)    +\  Q_R^1(z)  \Psi_R^1 (x) +... \, ,\label{sep2}\\
u_L(x,z)   &=&  u^0_L(z)  q_L^0(x)     +\  U_L^1(z)  \Psi_L^1 (x)+...\,  ,\label{sep3}\\
{u}_R(x,z) &=&  u^0_R (z) {u}_R^0(x)   +\  U_R^1(z)  \Psi^1_R(x) +...\, ,\label{sep4}
\eea
where $q^0_L(x)$ and $u^0_L(x)$ are the SM fermions and $\Psi^n_{L,R}(x)$ are the heavier
KK modes. In order to obtain a chiral spectrum, we choose 
boundary conditions for the fermion wavefunctions
\begin{eqnarray}
q_L (+ +),\quad q_R(- -), \quad u_L(- -), \quad u_R(+ +),
\end{eqnarray}
so that before electroweak symmetry breaking only $q^0_L$ and $u^0_R$
will be massless (zero modes) with wavefunctions: 
\begin{eqnarray}
q_L^0(z) &=& f(c_q)\frac{{R'}^{-\frac{1}{2}+c_q}}{ R^{2}} z^{2-c_q}, \\
u_R^0(z) &=& f(-c_u)\frac{{R'}^{-\frac{1}{2}-c_u}}{ R^{2}} z^{2+c_u},
\end{eqnarray}
where we have defined $f(c) \equiv
\sqrt{\frac{1-2c}{1-\epsilon^{1-2c}}}$ and the hierarchically small
parameter $\epsilon=R/ R'\approx 10^{-15}$. Thus, if we choose $c_q
(-c_u) > 1/2$, the zero mode wavefunctions are localized towards
the UV brane; if $c_q (-c_u) < 1/2$, they are localized towards the IR brane.

In order to implement minimally the Higgs sector out of a 5D scalar we use the
following action \cite{gaugephobic} 
\begin{equation}
{\cal S}_{\text{Higgs}} = \int dz d^4x \left(\frac{R}{z}\right)^3
\left[ Tr |{\cal{D}}_M H|^2 - \frac{\mu^2}{z^2} Tr|H|^2  \right] -
V_{UV}(H)\delta(z-R) - V_{IR}(H)\delta (z-R'), 
\end{equation}
where $\mu$ is the 5D mass for the Higgs boson. The boundary
potentials $V_{UV}(H)$ and $V_{IR}(H)$ yield boundary conditions
that can accommodate electroweak symmetry breaking, so that one obtains
a Higgs VEV with a non-trivial profile along the
extra-dimension. Around that VEV, one should then add perturbations and
obtain the spectrum of physical modes, i.e. a SM-like Higgs boson and a
tower of KK Higgs fields. The expansion should look like
\bea
H(x,z)= v_\beta(z) + h_\beta(z) h(x) + ....,
\eea
and we can choose the boundary conditions
such that the profile of the Higgs VEV $v_\beta(z)$ takes the simple form
\begin{equation}
v_\beta(z)=V(\beta)\ z^{2+\beta} \label{vz},
\end{equation}
where $\beta = \sqrt{4+\mu^2}$ and
\begin{eqnarray}\label{vevprof}
V(\beta) = \sqrt{\frac{2(1+\beta)}{R^3 (1-
(R'/R)^{2+2\beta})}}\frac{v_4}{(R')^{1+\beta}},
\end{eqnarray}
where $v_4$ is the SM Higgs boson VEV.
One should note that the wave function $h_\beta(z)$ of the light physical
Higgs (lightest KK Higgs field) will have the form 
\bea 
h_\beta(z)=\frac{v_\beta(z)}{v_4} \left(1+{\cal
O}\left(\frac{ m_h^2 {z}^2}{1+\beta}\right) \right), 
\label{hvsv}
\eea 
so that for a light enough Higgs boson mass both profiles $h_\beta(z)$
and $v_\beta(z)$ are aligned (i.e. proportional to each other).

The previous bulk Higgs sector is capable of reproducing the brane Higgs limit,
since the wavefunction of the light Higgs (and its VEV) both depend exponentially on the
parameter $\beta$. As this parameter is increased, the wavefunctions
are pushed more and more towards the IR brane mimicking a perfectly
localized Higgs sector.\footnote{Moreover the masses of the heavier KK Higgs fields 
depend linearly on the $\beta$ parameter and so these fields will
decouple from the theory for very large $\beta$.}
Indeed, the wave function of the Higgs can act as a brane
localizer since 
\bea
\lim_{\beta\to\infty} h^2(z) =\lim_{\beta\to\infty} v^2(z) = \delta(z-R'),
\eea
where the Dirac delta function is defined as the limit of a sequence
of functions with increasing value of $\beta$. One can easily prove
that for any wavefunction $f(z)$ (or a product of wavefunctions) we have
\bea
\lim_{\beta\to \infty} \int_{R}^{{R'}^+}h^2(z) f(z) \, dz = f(R'). \ 
\eea
There is however an issue about localizing the whole Higgs sector
towards the brane since we just showed that only quadratic Higgs
operators will ``become'' brane localizers. When a 5D action
operator contains more than two (or less than two) Higgs fields, the (successful)
localization of such operators is not guaranteed. In fact in order to
ensure that the 5D bulk Higgs scenario correctly tends smoothly to a
fully localized Higgs sector, one should implement a prescription
enforcing a precise $\beta$ dependence on the coefficients of all
operators containing Higgs fields. More precisely, the coefficient
$Y^{N}(\beta)$ of an operator containing $N$ Higgs fields (before
electroweak symmetry breaking) should behave
as \bea
Y^N(\beta)= Y^N_1\times \beta^{\frac{2-N}{2}},
\eea where $Y^N_1=Y^N(1)$. 
This is the only way to ensure that we can have
\bea
\lim_{\beta\to \infty} \int_{R}^{{R'}^+} Y^N(\beta)\ h^N(z) f(z)
\, dz =\lim_{\beta\to \infty} \int_{R}^{{R'}^+} Y_1^N\ \beta^{\frac{2-N}{2}}h^N(z) f(z)
\, dz= Y^N_1\ f(R'), \ 
\eea
or in other words
\bea
\lim_{\beta\to\infty}  Y^N(\beta)\ h^N(z) = Y^N_1\ \delta(z-R').
\eea
In particular for 5D Yukawa type couplings this prescription implies
that the 5D Yukawa coupling will have to carry a $\sqrt{\beta}$
dependence in order to ensure that the brane limit Yukawa coupling is
non-vanishing \cite{Agashe:2006iy} (see also \cite{ATZ1}). But it also means that any
other 5D action operator containing a single Higgs field would need to
carry the same $\sqrt{\beta}$ dependence.  On the other hand, 5D
action operators containing  3 Higgs fields (like the operator
$H^2HQU$) would have a diverging limit for $\beta$ large unless its
action coefficient $Y^{3}(\beta)$ is itself suppressed by 
$1/\sqrt{\beta}$.

The previous prescription makes it technically possible to define a
localized Higgs sector from a 5D bulk Higgs field, but it certainly
seems quite contrived to appropriately fix all operator coefficients such
that they all can give non-zero and finite contributions when the Higgs is
localized. A brane localized Higgs sector could seem ``un-generic'' or
``un-natural'' if it is to be seen as a limiting case of a bulk
Higgs. More details about the complete prescription for  operators
containing Higgs fields are presented in Appendix \ref{apen:II}.

\section{Higgs phenomenology: all KK fermions}
\label{sec:infinite}

For completeness and consistency, we present first a result previously obtained in
\cite{ATZ1}, namely the computation of the shift between the light SM
fermion mass term and its Yukawa coupling with the Higgs field (leading to Higgs
mediated FCNC when more than one
fermion family is considered). We then calculate the coupling between the
physical Higgs and two gluons for the 5D bulk Higgs case.\footnote{We
   follow very closely the general procedure 
  outlined in \cite{ATZ2} and explicitly compute the prediction for
  the bulk Higgs case and then compare it with the brane localized Higgs
  limit that was presented there. For the sake of simplicity here
  we assume the matter fields belong to the usual SM gauge group.} 

We can follow two routes to obtain our predictions. The computation of
the flavor violating couplings of the Higgs scalar with
fermions will be obtained in an approach based on considering {\it first}
electroweak symmetry breaking and {\it then} solving the 5D
equations of motion for the fermions (i.e. the effect of the Higgs VEV
is directly taken into account in the equations of motion and during the
dimensional reduction procedure). The alternative (and equivalent) approach
would be to consider {\it first} the dimensional reduction  (i.e. obtain 
the 4D effective theory in the gauge basis), and {\it  then}  consider the 
electroweak symmetry breaking in the presence of the infinite tower of
KK fermions. After performing the diagonalization of the infinite
fermion mass matrix (as well as canonical normalization of the fermion
kinetic terms) we should recover the same results.  We use the first
approach in the first subsection, and the second approach in the
computation of the Higgs coupling to gluons and also in the following
sections where we will truncate the infinite mass matrix in order to
consider only the effect of the first few KK modes.

\subsection{Higgs Flavor violating couplings}
\label{subsec:infiniteFCNC}

After imposing electroweak symmetry breaking in the Higgs sector, the four
profiles $q_{L,R}(z)$ and $u_{L,R}(z)$ introduced in eqs.~(\ref{sep1})
to (\ref{sep4}) must obey the coupled equations coming from the equations of motion:
\bea
&& -m_u\ q_L - q'_R + {c_{q}+2\over z} q_R + \left({R\over  z}\right)v_\beta(z) Y_u\ u_R=0, \label{qr}\\
&& -m^*_u\ q_R + q'_L + {c_{q}-2\over z} q_L + \left({R\over z}\right)v_\beta(z) Y_u\ u_L=0, \label{ql}\\
&& -m_u\ u_L - u'_R + {c_{u}+2\over z} u_R + \left({R\over z}\right)v_\beta(z) Y^*_u\ q_R=0, \label{dr}\\
&& -m^*_u\ u_R + u'_L + {c_{u}-2\over z} u_L + \left({R\over
z}\right) v_\beta(z) Y^*_u\ q_L=0, \label{dl} \label{eqmot} \eea
 where the $'$ denotes derivative with respect to the extra coordinate $z$ and
$Y_u$ is 5D Yukawa coupling. 

It is simple to deduce from these equations an exact expression for the mass eigenvalue
$m_u$ in terms of the fermion profiles \cite{ATZ1}
\bea 
m_u=R^4\int^{R'}_R\!\!
dz \left(\frac{m_u}{z^4} (|u_L|^2+|q_R|^2) + {Rv_\beta(z)\over z^5}(Y_u
u_Rq^*_L\!-\!Y^*_uq_Ru^*_L)\right),  \ \ 
\eea 
and compare it to the expression of the fermion Yukawa coupling, i.e
\bea y^u_4 = R^5\int^{R'}_R dz {h_\beta(z)\over z^5}(Y_u
u_Rq_L^*+ Y^*_dq_Ru_L^*), 
\eea 
where $h_\beta(z)$ is the profile of the physical Higgs field. 

\begin{figure}[t]
\center
\begin{center}
\includegraphics[height=8cm]{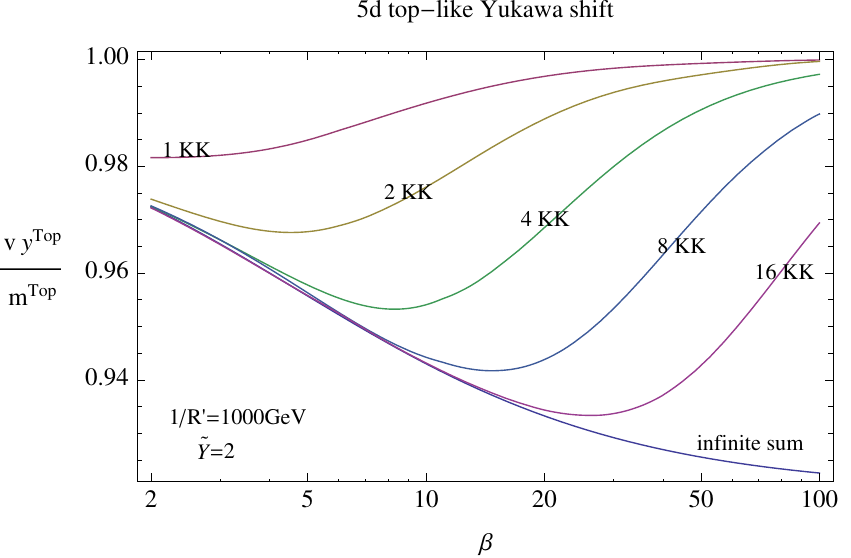}
\caption{(color online). The shift in the top quark Yukawa coupling as a
  function of the bulk Higgs localization parameter 
  $\beta$. Each line represents an effective theory containing the
  given amount of KK fermions. The lower line (blue) represents the
  contribution from the infinite tower of KK modes. Apart from the
  direct phenomenological impact of this result, this term also
  affects the $hgg$ coupling, as discussed in the text. The
  dimensionless 5D Yukawa couplings are fixed at $\tilde{Y}=2$ and the KK scale
  is set at $\displaystyle \frac{1}{R'}=1000$ GeV (the overall effect
  scales as $\tilde{Y}^2 v^2 R'^2$).}   \label{fig:topshift}
 \end{center}
\end{figure}

With these two expressions we compute the shift (or
misalignment) between the fermion mass $m_u$ and the Yukawa coupling $y^u_4$ as
\bea
\Delta^u=m_u - v_4\ y^u_4,
\eea 
which becomes simply
\bea
\Delta^u&=&R^4\int^{R'}_R\!\! dz
\left(\frac{m_u}{z^4} (|u_L|^2+|q_R|^2) -2Y^*_u {R v_\beta(z)\over z^5}
q_Ru_L^*\right). \
\label{Delta}
\eea 
In order to proceed further, a perturbative approach is used, such that we
assume that $(\tilde{Y}_uv_4R')\ll 1$ where $v_4$ is the SM Higgs
VEV. Knowing the analytical form of the VEV profile $v_\beta(z)$ and using
the $(\tilde{Y}_uv_4R')$ small parameter it is possible to solve
perturbatively the system of coupled 
equations (\ref{qr}) to (\ref{dl}) to any order in
$(\tilde{Y}_uv_4R')$ (see \cite{ATZ1} for details). The
result for the shift in the top quark Yukawa coupling is
\begin{eqnarray}
\frac{\Delta^t_{1}}{m{v_4}}&&=
\frac{2 m_t^2}{{v_4}} R'^2 \frac{2+c_u -c_q+\beta}{(1-2c_q)(1+2c_u)} 
\left [\frac{1}{6+c_u-c_q+3\beta}
-\frac{1}{5+2c_u+2\beta}
\right.
\nonumber\\
&&\left.
\hspace{1cm}-\ \frac{1}{5-2c_q+2\beta} + \frac{1}{4+c_u-c_q+\beta}
\right ], 
\end{eqnarray}
where we have only included the contribution from the third term
in eq.~(\ref{Delta}), as the other terms are subdominant for light
quarks, although not necesarily for the top quark. For
clarity we omit their analytical expression here, but    
the complete analytical result can be found in \cite{ATZ1} and in the
Appendix \ref{apen:I} of this work. The shift in the Yukawa 
coupling has some dependence on the Higgs localization parameter
$\beta$ and it is shown in Figure \ref{fig:topshift} as the ``infinite sum'' result,
as the  procedure we followed is equivalent to diagonalizing the infinite
fermion mass matrix in the gauge eigenbasis.

\subsection{Higgs production}
\label{subsec:infiniteprod}

In this section we follow the approach of working with the
infinite fermion KK modes with wavefunctions in the gauge basis. This
is not the physical basis after electroweak symmetry breaking since
Yukawa couplings will introduce off-diagonal terms in the infinite fermion mass
matrix, which should be properly diagonalized in order to obtain the physical basis.

\begin{figure}[t]
\center
\begin{center}

	\includegraphics[height=4cm]{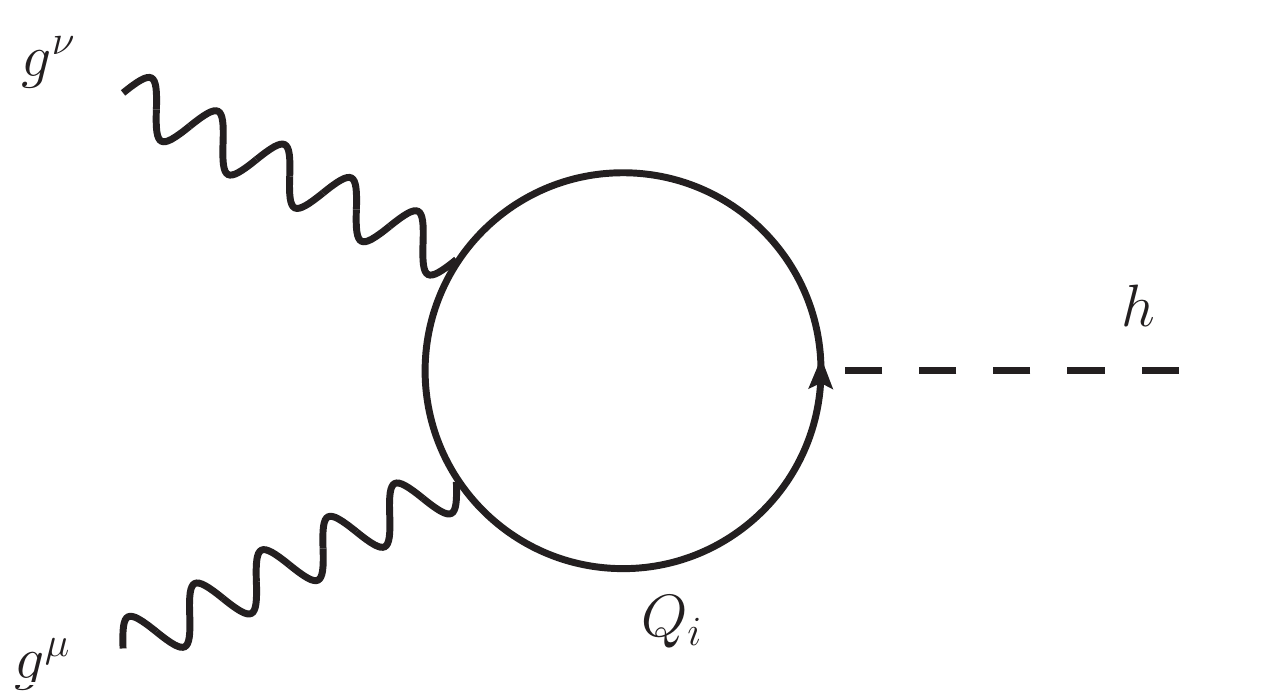}
 \end{center}
\caption{Loop diagram showing the contribution of the quark $Q_i$ to the
  Higgs-gluon-gluon coupling. In the SM, the dominant contribution is
  through the top quark due to its large Yukawa coupling with the
  Higgs boson. In RS the heavier KK fermions contribute to the coupling
  with potentially large effects, either suppressing or enhancing the
  SM coupling, depending on the phases present in the different
  Yukawa-type operators present in the 5D action, and on the
  localization of the Higgs (see text for details).} 
\label{fig:toploop}
\end{figure}

Since the Higgs field is not charged under QCD the main
contribution to its coupling to gluons comes from a top quark loop, as shown in
Figure \ref{fig:toploop}; if the model contains many heavy quarks
the resulting cross section for the process is
$gg\to h$ is \cite{higgs}   
\bea
\sigma^{SM}_{gg\rightarrow h} = {\alpha_s m_h^2\over
  576\pi}\left|\sum_Q{y_Q\over m_Q}
A_{1/2}(\tau_Q)\right|^2\delta({\hat s}-m_h^2)\label{hggSM},  
\eea
with $\tau_Q\equiv m^2_h/4m_Q^2$, $\hat{s}$ being the $gg$ invariant
mass squared and $Q$ representing the physical fermions with physical Yukawa couplings
$Y_Q$ and masses $m_Q$. The form factor is given by  
\bea
A_{1/2}(\tau) = {3\over 2}[\tau +(\tau -1)f(\tau)]\tau^{-2}\label{ff},
\eea
with
\bea
f(\tau) = \begin{cases}
      [\text{arcsin}\sqrt{\tau}]^2  \qquad\;\qquad\;\qquad{\tau\leq 0}\\
      -{1\over 4}\left[\text{ln}\left({1+\sqrt{1-\tau^{-1}}\over
          1-\sqrt{1-\tau^{-1}}}\right)-i\pi\right]^2
      {\tau>1}\label{ftau}. 
          \end{cases}
\eea   
Here we want to figure out the contribution to the $hgg$ coupling coming
from 5D quark doublets and a singlets, i.e. containing the SM quarks (which
includes doublets and singlets $(q_L,u_R)$), along with the associated
towers of vector-like KK fermions, $(Q_L, U_R)$.
The relevant quantity to calculate is 
\bea 
c_{hgg}=\sum_Q{y_Q\over m_Q} A_{1/2}(\tau_Q),
\eea 
where $y_Q$ is the physical Yukawa coupling of the physical Dirac fermion $Q$ and $m_Q$ is its
mass. As stated before, it will prove useful to work in the gauge basis, and so we 
represent the Yukawa couplings between the KK fermions $Q_L(x)$ and 
$U_R(x)$ in the gauge basis as $Y_{Q_LU_R}$. Its values will be
obtained by performing the overlap integral of the Higgs profile and
the corresponding bulk fermionic wave functions, i.e.  
\bea 
Y_{Q_LU_R}^u = Y^u\int_R^{R'}dz\left({R\over z}\right)^5
\frac{v_\beta(z)}{v_4} Q_L^{u(i)}(z)U_R^{(k)}(z)\label{yu}, 
\eea
where we have assumed that the nontrivial Higgs VEV and the physical
Higgs profile are perfectly aligned\footnote{We address the case where 
$h_\beta(z)\neq v_\beta(z)/v_4$ in Section \ref{sec:vevmisalignment}.}.
The Yukawa couplings between different chirality KK fermions and also 
between zero modes and heavy KK fermions are obtained and written in a
similar way so that we can write the infinite dimensional fermion mass matrix as
\bea
\begin{pmatrix}
            \bar{q}_L^{u(0)}& \bar{Q}_L^{u(i)}& \bar{U}_L^{(j)}
\end{pmatrix}\label{decom}
\begin{pmatrix}
   Y^u_{{q}_Lu_R}\  v_4       &          0            &   Y^u_{{q}_LU_R}\ v_4\\ 
   Y^u_{{Q}_Lu_R}\  v_4       &          M_Q          &   Y^u_{{Q}_LU_R}\ v_4\\ 
        0                  &  Y^{u*}_{{U}_LQ_R}\ v_4   &      M_U 
             \end{pmatrix}
\begin{pmatrix}
 u_R^{(0)}\\ Q_R^{u(k)}\\ U_R^{(l)} \label{infiniteM}
\end{pmatrix},
\eea 
where $M_Q = \text{diag}(M_{Q_1}, M_{Q_2}, ...)$ and  $M_U =
\text{diag}(M_{U_1}, M_{U_2}, ...)$ are the KK mass matrices for the corresponding 
fermion fields in the gauge basis, and we have suppressed fermion family
indices to simplify notation. From eqs.~(\ref{ff}) and (\ref{ftau}), we notice
that in eq.~(\ref{hggSM}) the form factors, $A_{1/2} \approx 0$ 
for light fermions, and $A_{1/2} \approx 1$ for the much 
heavier KK modes and the top quark. Therefore, separating the 
contribution of the light fermions from the heavy ones 
we write 
\bea
c_{hgg}=
\sum_{\rm light}{y_Q\over
  m_Q}A_{1/2}(\tau_Q) + \sum_{\rm heavy}{Y_Q\over M_Q},  
\eea
where in the first (second) term the sum
is only over light (heavy) fermion generations. Noting that 
\bea
\sum_{\rm heavy}{Y_Q\over M_{Q}} +  \sum_{\rm light}{y_Q\over m_Q} = \text{Tr}(\mathbf{YM}^{-1}) ,
\eea
where $\mathbf{M}$ is the fermion mass matrix given in 
(\ref{decom}), while $\mathbf{Y}$ is the Yukawa matrix, we have
\bea
c_{hgg}=
\text{Tr}(\mathbf{YM}^{-1}) + \sum_{\rm light}{y_Q\over
  m_Q}(A_{1/2}(\tau_Q) - 1)\label{sumym}. 
\eea
We also note that $Y={\partial M\over\partial{v_4}}$ and since the
trace is invariant under unitary transformations, we can compute it in
the gauge basis (so we can use the fermion mass matrix in that
basis). Up to first order in ${v_4}$ one finds 
\bea
\text{Tr}(\mathbf{YM}^{-1}) = {\partial\text{ln Det}(\mathbf{M})\over\partial{v_4}} \approx 
{1\over{v_4}}-{v_4}\sum_{i,j}{2\over M_{Q_i}M_{U_j}}\left(Y^u_{{Q_L}_i{U_R}_j}Y^{u*}_{{U_L}_j{Q_R}_i} - 
{Y^u_{q_L{U_R}_j}Y^{u*}_{{U_L}_j{Q_R}_i}Y^u_{{Q_L}_iu_R}\over
  Y^u_{q_Lu_R}}\right)\label{trace}.\ \ \ \ \ \
\eea
Noting that the SM masses and Yukawa couplings 
are also modified (shifted) as  \cite{ATZ1} 
\bea\label{ym}
{y_Q\over m_Q}\big|_{\text{light}} \approx 
{1\over{v_4}}\left(1 + 2{{v_4}^2\over
  Y^u_{q_Lu_R}}\sum_{i,j}{Y^u_{q_L{U_R}_j}Y^{u*}_{{U_L}_j{Q_R}_i}Y^u_{{Q_L}_iu_R}\over
  M_{Q_i}M_{U_j}}\right),    
\label{shiftsum}
\eea
we can write the total $hgg$ coupling as
\bea
c_{hgg}=-2{v_4}\sum_{i,j}{Y^u_{{Q_L}_i{U_R}_j}Y^{u*}_{{U_L}_j{Q_R}_i}\over
  M_{Q_i}M_{U_j}} + {y_Q\over  m_Q}\big|_{\text{light}}A_{1/2}(\tau_{Q_{\rm light}})
\label{C}. 
\eea
where we have used equations (\ref{sumym}), (\ref{trace}) 
and (\ref{ym}).
As we mentioned before, the form factor is negligible 
for the light fermion generations. Therefore neglecting 
the last term above, and using (\ref{yu})
we have
\bea\label{hggyu}
c_{hgg}=-2{v_4}Y^uY^{u*}R\sum_{i,j}\int dzdz'\left({R\over z}\right)^5\left({R\over z'}\right)^5
{Q^{(i)}_L(z)Q^{(i)}_R(z')\over
  M_{Q_i}}{U^{(j)}_R(z)U^{(j)}_L(z')\over M_{U_j}} h_\beta(z)h(z'),\ \ \ \ 
\eea
where the 5D bulk physical Higgs profiles can be 
normalized as \cite{gaugephobic}
\bea
h_\beta(z) = \sqrt{{2(1+\beta)\over R^3(1-\epsilon^{2+2\beta})}}R'\left({z\over R'}\right)^{2+\beta},
\eea
with $\epsilon\equiv R/R'\sim 10^{-15}$ being the warp factor. The sums in eq. (\ref{C})  are given by
\cite{ATZ2}
\bea
\sum_{i=1}^{\infty} 
{Q^{(i)}_L(z)Q^{(i)}_R(z')\over M_{Q_i}}
= -{z'^{2+c_q}z^{2-c_q} \over R^4}\left[\theta(z'-z)-{(z'/R)^{1-2c_q}-1\over\epsilon^{2c_q-1}-1}\right], \label{sumDbrane}
\eea
and
\bea
\sum_{j=1}^{\infty}
{U^{(j)}_R(z)U^{(j)}_L(z')\over M_{U_j}} 
={z^{2+c_u}z'^{2-c_u} \over R^4}\left[\theta(z'-z)-{(z'/R)^{1+2c_u}-1\over\epsilon^{-2c_u-1}-1}\right]\label{sumUbrane}.
\eea
Substitution of these sums and of the Higgs profile in Eq.~(\ref{hggyu}) and
assuming \footnote{For a completely flat bulk Higgs, 
   $\beta=2$. For any physically acceptable model 
  $\beta>2$.} $\beta\geq 2$ will finally give the total Higgs  coupling for the
light fermions which is given in Appendix I.  If we assume that
$c_q>1/2$ and $c_u<-1/2$, which is the case for light fermions
(up-like fermion), the expression for $c_{hgg}$ can be simplified as
\bea
c^{Up}_{hgg}\approx{v_4}Y^{u}Y^{u*}R'^2{2(1+\beta)\over(2+\beta+c_q-c_u)}{1\over4+2\beta}\label{4+2beta}.
\eea
In the case of the top quark, we have to add the contribution due to the last term
in Eq.~(\ref{C}), since $A_{1/2}(\tau_{top})\sim 1$. Following the notation in \cite{ATZ1}, 
we write the additional contribution as
\bea
{y_Q\over m_Q}\big|_{\text{light}}A_{1/2}(\tau_{Q_{light}}) + {\Delta_2^{top}\over m_t{v_4}},
\eea
where the first term is given by eq.~(\ref{ym}) multiplied by the
form factor, $A_{1/2}$ and last term, is the result of kinetic term
corrections due to the shift in Yukawa couplings, which are also not
negligible for the heavy fermions. The shift is given by 
\bea
\Delta_2^{top}&=&R^4\int^{R'}_R\!\! dz
\left(\frac{m_t}{z^4} (|u_L|^2+|q_R|^2)\right). \
\label{Delta}
\eea 
For a complete discussion on this, we refer the reader to \cite{ATZ1}. 

\begin{figure}[t]
\center
\begin{center}
\hspace*{-0.4cm}
	\includegraphics[width=3.5in]{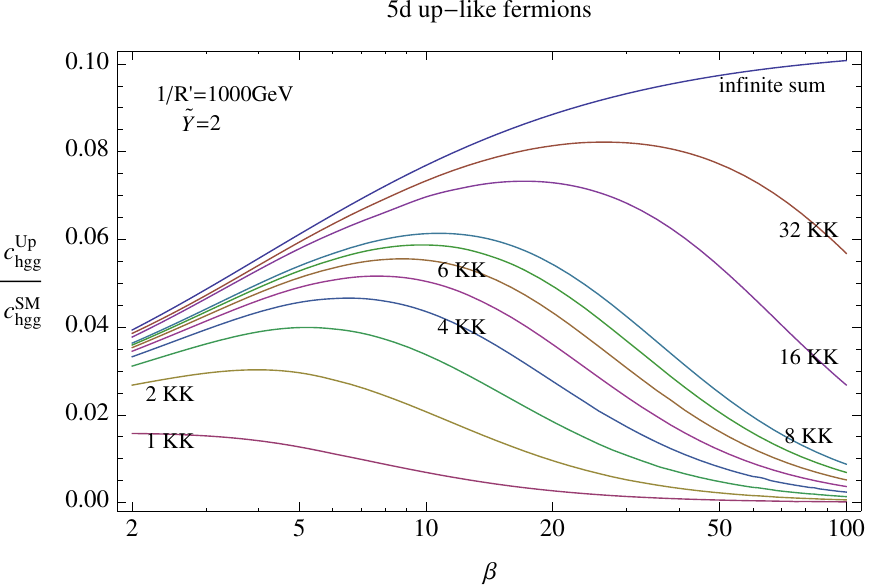}\includegraphics[width=3.5in]{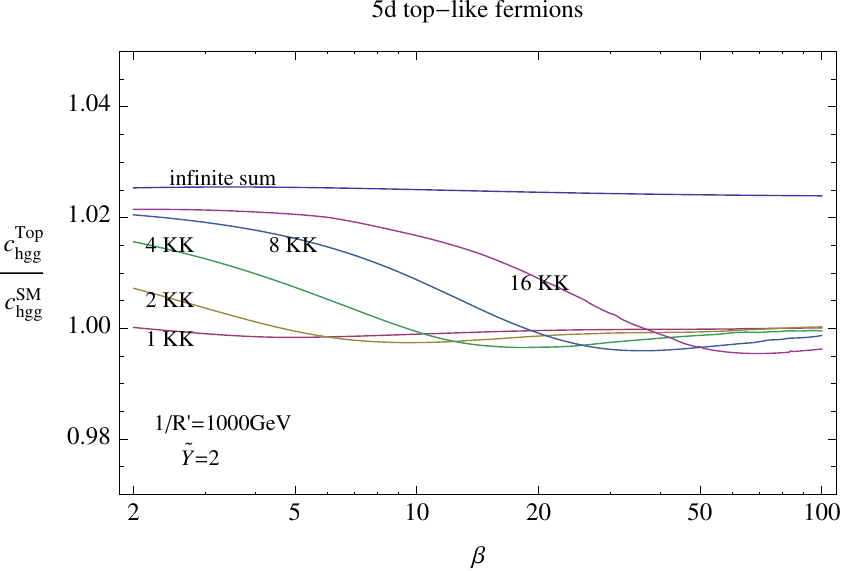}
 \end{center}
\caption{Contribution to $c_{hgg}/c_{hgg}^{SM}$ coming from the KK
  partners of the ``up'' quark (left panel) and from the full top quark
  sector (right panel) as a function of the bulk Higgs
  localization parameter $\beta$. Each line represents the numerical
  result obtained in an effective theory containing the amount
  of KK fermions indicated. The upper line (blue) represents the
  contribution of the infinite tower of KK modes (computed in the text
  analytically). The dimensionless 5D Yukawas are fixed at $\tilde{Y}=2$
and the KK scale is set at $\displaystyle \frac{1}{R'}=1000$ GeV (the overall
effect scales as $\tilde{Y}^2 v^2 R'^2$).} 
\label{fig:Cggh}
\end{figure}

So finally for IR localized fermions with $c_q<1/2$ and $c_u>-1/2$
(top-like) we have
\begin{eqnarray}
c^{Top}_{hgg}&\approx&{y_Q\over m_Q}\big|_{\text{light}}A_{1/2}(\tau_{Q_{light}}) + {\Delta_2^{top}\over m_t{v_4}}\nonumber\\
&&\
-{v_4}Y^{u}Y^{u*}R'^2\left[-{1\over4+2\beta}+{1\over 2\beta
    +5-2c_q}+{1\over 2\beta +5+2c_u}-{1\over\beta+4-c_q+c_u}\right]. 
\end{eqnarray}
Following our ansatz for localizing the Higgs sector, and in order to
compare with previous brane Higgs results, we need 
to replace the $5D$ Yukawa couplings with the dimensionless and $\beta$-independent couplings
\bea
\tilde{Y} = {\sqrt{2(1+\beta)}\over(2-c_q+c_u+\beta)}Y^{5D}.
\eea
The results obtained in this section, of the contribution of a 5D top-like quark and a 5D
up-like quark to the $hgg$ coupling are shown in both panels of Figure
\ref{fig:Cggh} as the ``infinite sum" result.

\section{Higgs phenomenology: individual KK modes}
\label{sec:individual}

In this section we take a different approach and compute the effects on
Higgs phenomenology (FCNC and production cross section) due to only
the first few KK fermions in the model. That is, we consider a
4-dimensional effective theory which contains the SM matter content,
augmented by a few levels of KK fields. This procedure is better
fitted within the framework we work in (low cut-off effective
theories), the drawback being that it is not possible to obtain
general analytical predictions in a close form. Our strategy will be
to assign some generic values to the parameters of the model and
perform the computations numerically. In particular we will fix the
bulk mass parameters of the 5D fermions $Q$ and $U$ to be $c_u=-0.6$
and $c_q=0.6$ (for an up-type quark) and $c_u=0$ and $c_q=0.4$ (for a
top quark). The value of the dimensionless 5D Yukawa coupling will be
taken to be $\tilde{Y}^{5D}_1=2$.

\subsection{Higgs Flavor violating couplings}
\label{subsec:individualFCNC}

In order to evaluate the shift in the Yukawa coupling of the SM
fermion (the zero mode) due to the presence of a finite number of KK
fermions, we can simply use Eq.~(\ref{shiftsum}), with the understanding that
now the sum is finite, and so we shall sum up to the maximum number of KK modes chosen.
We are interested in computing the top quark Yukawa shift as it is the
most interesting for direct phenomenology, and also because it will
also enter in the calculation of the $hgg$ coupling. We perform the
sum numerically and stop the summation at different maximum numbers of KK
fermions. The results are shown in Figure \ref{fig:topshift} in which
we focus on the variation of the Yukawa coupling shift with respect to
the bulk Higgs localization parameter $\beta$ and we compare these to
the results obtained in the previous section for the infinite KK
degrees of freedom. The main observation is that for small $\beta$,
the finite sums are in good agreement with the infinite sum result. On
the other hand for large values of $\beta$ the Yukawa shift obtained
from the finite sums becomes more and more irrelevant and is clearly
at odds with the infinite sum prediction.

\subsection{Higgs production}
\label{subsec:individualprod}

To evaluate the contribution to the $hgg$ coupling coming from the
individual KK fermion modes we proceed as in the previous
subsection. We now use Eq.~(\ref{hggyu}), and sum up to the maximum
number of KK modes desired. 
We perform the sum numerically and show the results in Figure
\ref{fig:Cggh}. Again we are interested in the variation of the couplings
with $\beta$ and compare them to the result for the $c_{hgg}$
obtained by calculating the infinite sum, as shown in the previous
section.  

The two panels of the figure show the contribution to the
$hgg$ coupling coming from a 5D up-like quark (left panel) and the
contribution coming from a 5D top-like quark (right panel) for $\beta$
values up to 100. We can see how the sums over different maximum number of KK
modes converge to the infinite sum limit as we vary $\beta$. The  
approximation obtained by considering just a few KK modes is much
better for low values of $\beta$. For example, from the left panel of
Figure \ref{fig:Cggh}, for $\beta=2 \to 5$, 8 KK 
modes saturate some 90\% of the infinite sum, while for $\beta=20$, 8 KK
modes saturate some 60\% of the infinite sum. For $\beta=100$
(corresponding to a Higgs highly localized towards the brane),  8 KK 
modes represent only some 10\% of the total KK contribution. This
dependence on $\beta$ is in agreement with the results found in
\cite{Bouchart:2009vq}, in which a brane localized Higgs was
considered (i.e. $\beta=\infty$) and the first few KK fermions
considered were found to give a negligible contribution to the $hgg$
coupling. \\

We conclude that in all the previous calculations (the top quark
Yukawa shift and the contributions to the $hgg$ coupling coming from
up-type and top-like 5D quarks) we have observed the same feature,
namely that in the case of a bulk Higgs (small $\beta$), the effect of the heavier KK
modes decouples (i.e. performing the infinite sum is equivalent to sum
only over the first few KK modes). On the other hand, when $\beta$ is
very large, the heavier degrees of freedom do not seem to decouple
hinting towards some type of UV sensitivity of the brane Higgs case. 
This is not that surprising since the thickness of a Higgs being
crushed against the brane is becoming smaller and smaller, and the
scale associated with the Higgs localization eventually becomes much
larger than the cut-off of the scenario. We will now see how adding a
type of higher derivative operators will be sufficient to make the
finite sums consistent with the infinite sum results obtained earlier.

\section{The Effect of Higher Derivative Operators}
\label{sec:ramanop}

We have just seen how the results obtained in the previous section (\ref{sec:individual}),
where we sum over a few KK modes agree with the complete KK tower
summation of section \ref{sec:infinite} only in the case of a bulk
Higgs boson. When the Higgs is on the brane, or very much pushed
towards the brane, the results for the two approaches do not seem to agree (see Figure
\ref{fig:Cggh} when $\beta \to 100$).  We will reconcile the two
methods by including, in the effective theory 
calculation, the contribution of higher derivative operators.  

In particular we consider the effect of the following
operator in the action with a dimensionful coupling constant $Y_R$
(flavor indices are suppressed),
\bea
\hspace{-1cm} S \supset \int d^4x dz \sqrt{g}
\left[Y_{R}\ \overline{\Gamma^M {\cal
      D}_M\calQ} \ 
{\cal H} \Gamma^N {\cal D}_N \calU + h.c.\right].
\ \ \
\eea
The operator is of Yukawa-type as it couples two fermions with
the Higgs, but it involves derivatives of fields. The coupling $Y_R$
should be in units of $\Lambda$, the cut-off of the theory, and so obviously this operator 
is cut-off suppressed (we note that the standard 5D Yukawa $Y_u$ coupling is
also dimensionfull and cut-off suppressed, but by two units less than
$Y_R$). 
Since $Q_R(z)$ and $U_L(z)$ satisfy Dirichlet 
boundary conditions on the IR brane, their derivatives along the extra
dimension can be large after electroweak symmetry
breaking and so we focus on the operator  
\bea
\hspace{-1cm} S \supset \int d^4x dz \left({R\over z}\right)^3
\left[{Y_R}\overline{\partial_z q}_RH\partial_z
  u_L+h.c.\right],
  \label{R} 
\eea
which includes only the $wrong$ chirality fermion
components $Q_R(z)$ and $U_L(z)$ as it could lead to potentially
large effects. 

As explained in the previous sections we can proceed in two ways in
order to compute the effects of this operator. We could study the
effect of the operator into the 5D equations of motion {\it after}
electroweak symmetry breaking (ESB) and calculate its effects from
these. Alternatively, we could solve the equations of motion and
perform the dimensional reduction {\it before} ESB, and then consider the
effects produced by the operator by working in this gauge
eigenbasis. Both methods should be equivalent, but we will follow
the second one. In this approach, we obtain the effective 4D
theory and since it is non-renormalizable, we cut-off its spectrum at
the cut-off scale thus effectively we only allow a few physical
KK modes into the calculation. The effects from higher modes are integrated out and
encoded in all higher order operators of the theory with their effects
under control by the cut-off suppression.
In the case of the $Y_R$ operator the potentially large derivatives of $Q_R(z)$ and $U_L(z)$ can
offset the cut-off suppression and so we should keep this operator in
the calculations.
 
In the approach in which the KK modes are in the gauge basis,
the $Y_R$ operator will affect the fermion mass matrix from
eq.~(\ref{infiniteM}), and in particular it will contribute to the
$Y^u_{U_LQ_R}$ terms.
\begin{figure}[t] 
\includegraphics[height=8cm]{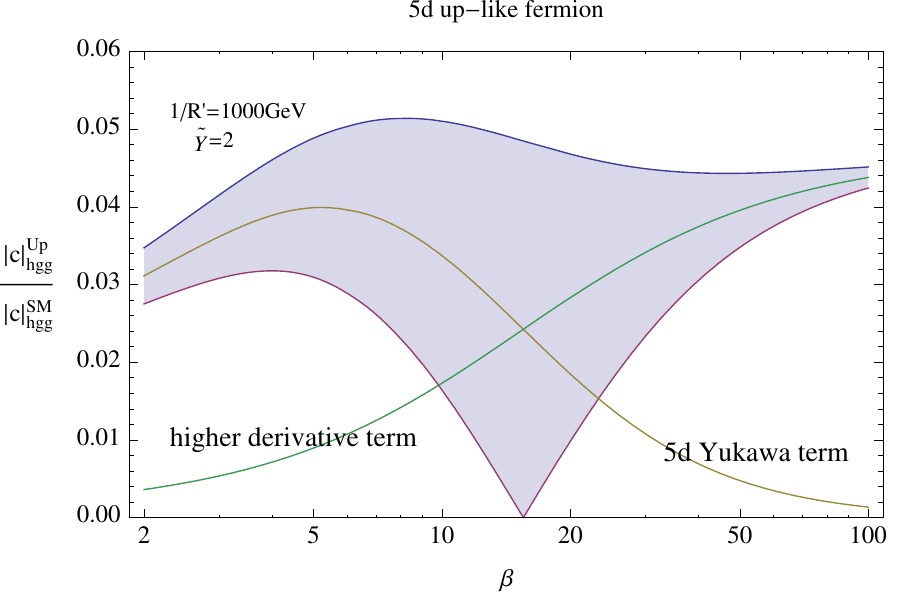}
\caption{(color online). Contribution to the coupling
  $|c_{hgg}|$ (relative to the Standard Model) as a function of the
  Higgs localization parameter $\beta$ when considering {\it only} a
  five-dimensional up-type quark, and 
  computed with the higher derivative term discussed in the
  text in addition to the standard 5D Yukawa coupling term. 
Since both contributions have independent phases we add and subtract
their generic size to obtain the shaded region of possible values.
These results are calculated by using only the first 3 KK modes
(i.e. considering an effective theory with a cut-off of the order the
the fourth KK mass). The dimensionless 5D Yukawas are fixed at $\tilde{Y}=2$
and the KK scale is set at $\displaystyle \frac{1}{R'}=1000$ GeV (the overall
effect scales as $\tilde{Y}^2 v^2 R'^2$).} 
\label{Rl}
\end{figure}
\begin{figure}[t] 
\center
\begin{center}$
\begin{array}{cc}
\hspace*{-1.5cm}
\includegraphics[width=3.5in,height=7cm]{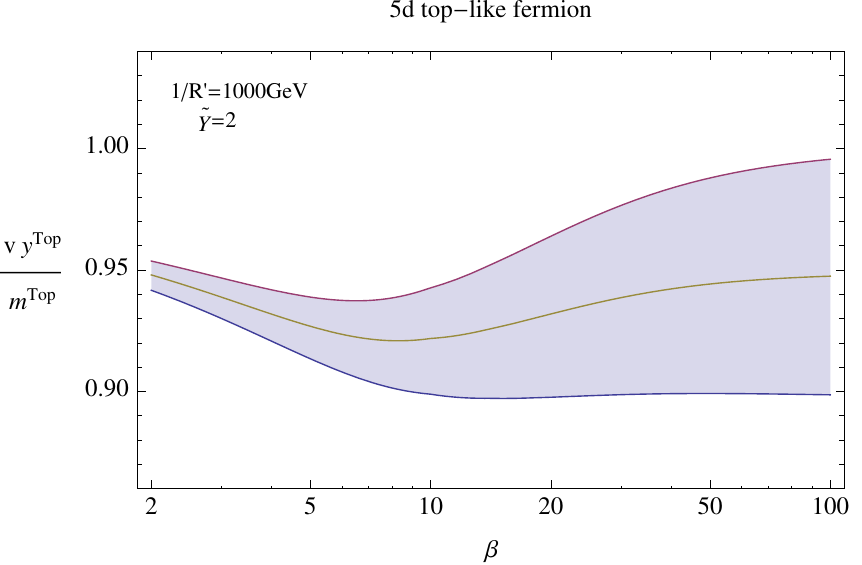}
&
\includegraphics[width=3.5in,height=7cm]{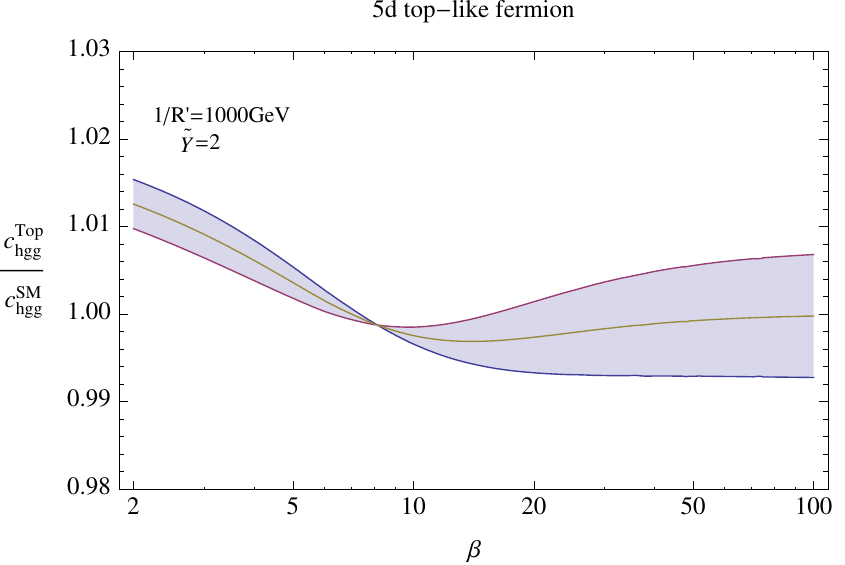}
\end{array}$
\caption{(color online). Shift of the top quark Yukawa coupling
  (left) and contribution to the coupling $c_{hgg}$ (right),
  relative to the Standard Model, as a function of the  
  Higgs localization parameter $\beta$, when considering only a
  five-dimensional KK top quark, and including in the computation the
  higher derivative term discussed in the text, in addition to the
  standard 5D Yukawa coupling term. The contributions from each term
  have independent phases and so we add and subtract their absolute
  value to obtain the shaded region of possible values. These results
  are calculated by using only the first 3 KK modes (i.e. considering
  an effective theory with a cut-off of the order the the fourth KK mass). The values of the 5D Yukawa $Y^u$ and of $Y_R$ are fixed at $\tilde{Y}=2$
and the KK scale is set at $\frac{1}{R'}=1000$ GeV (the overall
effect scales as $\tilde{Y}^2 v^2 R'^2$).} 
 \label{Rh}
 \end{center} 
\end{figure}
Its effects can therefore be tracked into the effects of these {\it wrong}
chirality terms, as was already noted in the appendix of \cite{ATZ1}.
We can thus formally treat the situation as before,
where a truncated version of the infinite mass matrix of eq.~(\ref{infiniteM})
is considered (with just a few KK levels), but now we redefine the
terms $Y^u_{U_LQ_R}$ to include the contributions from  $Y_R$ as 
\bea 
Y_{U_LQ_R}^u = \int_R^{R'}dz\left({R\over z}\right)^5
\frac{v_\beta(z)}{v_4} \left( Y^u U_L(z)Q_R(z) + Y_R^u {z^2\over R^2} \partial_zU_L(z)\partial_zQ_R(z) \right). 
\eea
It is now easy to compute numerically the new effects since from here we
just have to repeat the previous procedure. The results are shown in Figures
\ref{Rl} and \ref{Rh}. In both figures we show the individual
contributions coming from the normal Yukawa coupling $Y^u$, from
the new $Y_R^u$ coupling, as well as the combined effect. This
combined effect is represented by the shaded region, the reason being
that the two types of couplings $Y^u$ and $Y_R^u$ have independent
phases and so can add up constructively or destructively, or in between. 
In Figure \ref{Rl} we focus on the contribution to $hgg$ due to an
up-like 5D quark. In Figure \ref{Rh} we show the predictions for
the both shift in the SM top quark Yukawa coupling as well as the
prediction for the contribution to the $hgg$ coupling coming from a
 5D top-like quark. As we can see, the shift in the top quark Yukawa
coupling can be quite large, and for low values of the Higgs localization parameter $\beta$
the shift obtained  always results in a suppression in the Yukawa coupling. For large values of
$\beta$ the shift can be in either direction (suppression or enhancement).
In the case of the $hgg$ coupling, we see that the contribution represents
an enhancement with respect to the SM prediction for small values of
$\beta$, and again for large values of $\beta$ the $c_{hgg}$ coupling can be
either enhanced or suppressed depending on the relative phases between
$Y_R$ and $Y^u$. For this comparison we have taken the
absolute value of both couplings $Y_R$ and $Y^u$ to be the same, i.e. 
$\tilde{Y}=2$, in appropriate units of the cutoff. The main feature to
remember is that the effects of the higher derivative operator $Y_R$
are subdominant for small $\beta$ but become dominant for large $\beta$.
The large contribution obtained at large $\beta$ is precisely what
makes these new predictions consistent with the results obtained with
the original infinite sum, and so the higher derivative operators
that we have considered here somewhat encode the UV sensitivity found in the
previous section. 
\section{Misalignment between Higgs VEV and Higgs profile}
\label{sec:vevmisalignment}

In this section we present a discussion on how to treat the case where 
the Higgs profile is different form its VEV profile. This is
equivalent to consider the mixing effects between the massless zero
mode Higgs boson, and the heavy KK Higgs modes and its effects on the Higgs
observables computed in this paper.

We follow closely an argument by Azatov \cite{azatovhiggsvev} and for
simplicity we will discuss a simple situation in which the 4D
effective theory contains only two new heavy vector-like fermions, $Q$
and $U$, doublet and singlet of $SU(2)_L$ respectively. This is the
situation one would have when the KK fermion towers are truncated after
the first KK excitation.  

Let's first define our notation for the following quantities
\bea
Y^\beta_{ij}\equiv\int dz\left(\frac{R}{z}\right)^5 \psi_i\ \psi_j \frac{v_\beta(z)}{v_4}\nonumber\\
X^\beta_{ij}\equiv\int dz\left(\frac{R}{z}\right)^5 \psi_i\ \psi_j\ h_\beta(z),
\eea
where $v_4$ is the SM Higgs VEV and $v_\beta(z), h_\beta(z)$ are the 5D profiles
of the Higgs VEV and the Higgs 
physical field, which are generically different. That is, after EWSB,
the Higgs field is expanded around the nontrivial VEV $v_\beta(z)$ as 
\bea
H(x,z)=v_\beta(z) + h(x) h_\beta(z) + ...\, .
\eea
In the case of the bulk Higgs sector considered here, both profiles
$v_\beta(z), h_\beta(z)$ are almost the same, (see eq.~(\ref{hvsv})), the order
of the misalignment between them being controlled by powers of $(m_hR')^2$ (a small
quantity).

We consider all the possible couplings between the Higgs and the
fermions of the effective theory which after EWSB can be written as
the matrix ${\cal M}(v_4,h)$ as
\bea
\hspace{-1cm}\l(\bar{q_L},\bar{Q_L}\bar{U_L}\r)
\l(\baa{ccc}
Y^\beta_{q_Lu_R}\ v_4 + X^\beta_{q_Lu_R} h(x)      &      0          &  Y^\beta_{q_LU_R}\ v_4 + X^\beta_{q_LU_R}h(x)\\
Y^\beta_{Q_Lu_R}\ v_4 + X^\beta_{Q_Lu_R}h(x)       &     M_Q         &  Y^\beta_{Q_LU_R}\ v_4 + X^\beta_{Q_LU_R}h(x)\\ 
        0                    & Y^\beta_{U_LQ_R}\ v_4 + X^\beta_{U_LQ_R}h(x) &  M_U 
\eaa\r)
\l(\baa{c}u_R\\Q_R\\U_R\eaa\r),\ \ \ \ \ 
\eea
The coupling between the physical Higgs and the two gluons is controlled
by the physical Yukawa couplings $Y^{\rm phys}_{i}$ and masses $M^{\rm phys}_i$ of the heavier
physical fermions running in the loop (top quark and KK modes), i.e. 
\bea
\sum_{\rm heavy}\frac{Y^{\rm phys}_{i}}{M^{\rm phys}_i}=Tr({\bf Y}_{\rm phys}
{\bf M}_{\rm phys}^{-1})-\sum_{\rm light}\frac{y_{i}}{m_i},
\eea
where ${\bf Y}_{\rm phys}$ is the physical Yukawa coupling and ${\bf
  M}_{\rm phys}$ is the physical fermion mass matrix of the setup. 
Because the trace is invariant under unitary transformations, we can
rotate to the gauge basis and write
\bea
Tr({\bf Y}_{\rm phys}{\bf M}_{\rm phys}^{-1}) = Tr({\bf Y}_{\rm gauge}{\bf M}_{\rm gauge}^{-1}) ,
\eea
and note that we can now relate this to the matrix  ${\cal M}(v_4,h)$ as
\bea
Tr({\bf Y}_{\rm gauge}{\bf M}_{\rm gauge}^{-1}) = \d_h\log(Det {\cal M}(v_4,h))|_{h=0}.
\eea
The procedure is the same as was followed in Section
\ref{sec:infinite}, i.e. we compute the the determinant by expanding
in powers of $v^2/M_i^2$ and after combining everything we obtain 
\bea
\sum_{\rm heavy}\frac{Y_i^{\rm phys}}{M_i^{\rm phys}}&=&\d_h\log Det {\cal M}(v_4,h)-\frac{y^{\rm light}}{m^{\rm light}}\nonumber\\
&=&v_4\l(-\frac{X^\beta_{Q_LU_R}Y^\beta_{U_LQ_R} }{M_Q M_U}-\frac{X^\beta_{U_LQ_R}Y^\beta_{Q_LU_R}}{M_Q M_U} \r).
\eea
This result is the equivalent to eq.~(\ref{C}) with the effect of the
misalignment between $v_\beta(z)$ and $h_\beta(z)$. One sees that 
the difference lies in the substitution of one of the $Y$ terms by an
$X$ term, and so the correction to the result of eq.~(\ref{C}) is 
\bea
\delta{c_{hgg}}=v_4\l(-\frac{\left(X^\beta_{Q_LU_R}-Y^\beta_{Q_LU_R}\right)Y^\beta_{U_LQ_R} }{M_Q M_U}-\frac{\left(X^\beta_{U_LQ_R}-Y^\beta_{U_LQ_R}\right)Y^\beta_{Q_LU_R}}{M_Q M_U} \r),
\eea
which is controlled by 
\bea
\left(X^\beta_{U_LQ_R}-Y^\beta_{U_LQ_R}\right)= \int dz\left(\frac{R}{z}\right)^5
U_L(z)Q_R(z) \left(h_\beta(z)-\frac{v_\beta(z)}{v_4}\right),  \ \ \ \ \ 
\eea 
and
\bea
\left(X^\beta_{Q_LU_R}-Y^\beta_{Q_LU_R}\right)= \int dz\left(\frac{R}{z}\right)^5
Q_L(z)U_R(z) \left(h_\beta(z)-\frac{v_\beta(z)}{v_4}\right),  \ \ \ \ \ 
\eea 
and since the misalignment between $v_\beta(z)$ and $h_\beta(z)$ can be computed
perturbatively \cite{ATZ1} as
\bea
h_\beta(z)-\frac{v_\beta(z)}{v_4}=\frac{v_\beta(z)}{v_4} \left(\frac{m_h^2 {R'}^2}{2(4+\beta)}+ \frac{m_h^2
  {z}^2}{4(1+\beta)}+{\cal O}(m_h^4{R'}^4)\right),
\eea
we obtain
\bea
\left(X^\beta_{U_LQ_R}-Y^\beta_{U_LQ_R}\right) = m_h^2 {R'}^2
\left(\frac{Y^\beta_{U_LQ_R}}{2(4+\beta)} +
\frac{Y^{\beta+2}_{U_LQ_R}}{4\sqrt{(1+\beta)(3+\beta)}} \right)+{\cal O}(m_h^4{R'}^4).
\eea
In other words, the effect of considering the misalignment between $v_\beta(z)$
and $h_\beta(z)$ is to add a correction with the same structure as the
result of eq.~(\ref{C}), but with a suppression of $\left(m_h
{R'}\right)^2$, i.e. the correction is at most ${\cal  O}(1\%)$, and
becomes much smaller for increasing values of $\beta$.\footnote{The
  dependence on $\beta$ of the integrals $Y^\beta_{U_LQ_R}$ and
  $Y^\beta_{Q_LU_R}$ is quite mild and so, in terms of order of
  magnitude, we have $Y^\beta_{U_LQ_R} \sim Y^{\beta+2}_{U_LQ_R}$. }

\section{Discussion and Outlook}
\label{sec:conclusion}

In this work we have presented the results for the predictions of Higgs
phenomenology in a toy-model RS setup in which the Higgs field is allowed to propagate in the
bulk and with a single 5D fermion field. Our results can be
extended to three families to include full flavor effects, but the
generic predictions that we would obtain are expected to be basically
the same as the ones presented in \cite{ATZ1,ATZ2}. That is, that in the context of flavor 
anarchy, where the action parameters are all of the same order but
with more or less random values and phases (with the constraint of
obtaining correct SM predictions) the couplings of the Higgs with
fermions and gluons and photons can receive important corrections,
either enhancing or suppressing the SM predictions. 
However, the two references mentioned present calculations performed by
including the effect of all the KK fermions, technically assuming an
infinite cut-off for the model (where a brane Higgs is
considered). In general, all these scenarios break down at a low cut-off, becoming 
strongly coupled for both gauge and Yukawa interactions. The implicit
assumption made in \cite{ATZ1,ATZ2} was that the effects of the 
heavier modes should decouple quickly, at least for the case of a bulk
Higgs field. The main motivation to perform the calculations by considering
the full infinite fermion KK tower, as well as pushing the Higgs into the brane
was mainly of technical nature. Indeed both the flavor structure of
the Higgs Yukawa couplings as well as the coupling to gluons and
photons can be computed analytically with those ingredients. In
\cite{ATZ1}, the authors checked analytically that the corrections to
the Higgs Yukawa couplings were actually of the same order for a bulk Higgs and
a brane Higgs.

However it was pointed out in \cite{Neubert} that in the
brane Higgs case, the effects of the heavier KK fermion modes do not
decouple and that they all contribute evenly in the computation of the Higgs
couplings in the model. On the other hand, we showed in sections
 \ref{sec:infinite} and \ref{sec:individual} of this paper that the heavier KK
modes in the case of a bulk Higgs do decouple very quickly, so that the
analytical result obtained by using the infinite KK tower approaches
with great precision the numerical result obtained by considering an
effective theory with only a few KK fermion modes. Moreover, when
considering the effective theory with only a few KK modes, one should
include in the action all possible operators and in particular the
higher derivative ones introduced in Section \ref{sec:ramanop}. These
effects were omitted in \cite{Neubert}, and as we showed in this work, the
importance of these operators increases as the Higgs is more and  more
localized towards the brane. In \cite{Bouchart:2009vq}, the authors
considered an RS setup with a highly localized Higgs and the presence
of only a few KK fermions and studied the effects on the Higgs
couplings to gluons and photons, among other observables. In the limit
of the SM gauge group (they did consider an extended gauge group) they
found no significant deviations from the SM predictions. Indeed this
result is consistent with our findings of Section \ref{sec:individual} (no
higher derivative operators invoked yet), since as it can be seen on
Figures \ref{Rh} and \ref{Rl}, the shift in Higgs Yukawa couplings and
the new effects to  Higgs-gluon-gluon coupling vanish in the limit of
highly localized Higgs (large $\beta$ parameter). On the other hand,
in \cite{Neubert} it is claimed that large effects should be
present in the case of a brane Higgs and with only a few KK modes present in the
effective theory (and no higher derivative operators), a result
inconsistent with both our findings and those found in \cite{Bouchart:2009vq}.
We can trace the origin of the disagreement in their calculation of the Higgs
Yukawa couplings. Those are computed by using the full 5D equations of
motion, which as we have said earlier is equivalent to considering the
complete tower of KK modes. Then, using these couplings, they calculate
the $hgg$ radiative coupling but now including only a finite amount
of KK fermions. This treatment leads to a highly suppressed top quark Yukawa coupling (due to
 effects from the infinite KK tower) and a vanishing contribution to $hgg$ from the loops of KK
fermions considered (one would need the whole tower to obtain a 
finite effect). Their end result is a suppressed top quark Yukawa coupling and
a suppressed $hgg$ coupling (due to the smaller top quark Yukawa),
predictions which are at odds with the findings of
\cite{Bouchart:2009vq, ATZ2} and of this paper. 

The procedure of \cite{Neubert} seems inconsistent
because essentially the authors use infinite KK degrees of freedom
in one part of their calculation (the SM quark Yukawa couplings computation via
equations of motion) but then they truncate the KK degrees of freedom 
in order to compute the $hgg$ coupling. In any case, had they included
the higher derivative operators introduced in this paper, their
results would have changed dramatically since then, the effect to $hgg$
coming from the top quark Yukawa loop would remain basically the same, but the
effects due to loops of a few KK fermions would dominate the overall
effect (and thus the result would start to become consistent with the findings of \cite{ATZ2}).

Also,  the predictions of \cite{Bouchart:2009vq} should change if one
considers the effects of the higher derivative operators introduced in
Section \ref{sec:ramanop}. In that situation, the Higgs couplings can receive
large corrections, and can be of any sign (suppression or enhancement)
due to the different phases present in the couplings $Y$ and $Y_R$. 
In fact we have found here that for a Higgs field in the bulk, our results are more
predictive than for a brane Higgs field, because the effect of the higher
derivativer operators is subdominant for a bulk Higgs field.\footnote{Again,
  the reason for  
  this is that the value of the derivatives of the bulk fermions is
  suppressed by the higher value of the 5D cutoff. When the Higgs boson is
  pushed towards the brane, the derivatives of these fermions fields (with
  the ``wrong'' chirality) becomes larger and larger, and the 5D cutoff
  does not suppress anymore the effect of these operators.} 
The effects from only the 5D Yukawa operators are aligned \cite{ATZ1},
and thus all the KK quarks add up in phase. 
In that situation  we can have definite predictions for the effects
caused by a single family of fermions, i.e. it will produce a
suppression in the light quark Yukawa coupling and an enhancement in
Higgs boson production (as well as suppression in the Higgs to photons
coupling) \cite{ATZ2}, with the caveat of taking the
dimensionless couplings of both Yukawa terms and higher derivative
operators to be the same (consistent with the usual assumption that all
5D coefficients have to be of the same order). Taking into account the
three fermion families in conjunction with a bulk Higgs field might weaken this 
prediction due to complicated flavor mixings and structure, but still one should be
able to draw a correlation between Yukawa couplings and Higgs
production (and $h \to \gamma\gamma$) for the case of a bulk
Higgs field. The parameter space of the bulk Higgs scenario 
can therefore be under a tighter pressure as more and more precise
experimental measurements in Higgs observables at the LHC become
available. In particular if the predicted and correlated deviations of
Higgs couplings is not clearly observed this should put bounds on the
KK scale of the bulk Higgs scenario.

The situation for a Higgs on the brane is different. The higher order derivative
terms are now important. Each KK tower of light quarks and the top
will contribute to the $hgg$ coupling, but their effect depends on
arbitrary relative phases (between $Y_R$ and $Y_{5d}$), and so one
cannot make a firm statement about the magnitude and phase of the
overall contribution: it can be a suppression or an enhancement, or in
between. 

Finally we comment again on the apparent problem of a highly localized Higgs scenario
(brane Higgs) in which predictions made from a truncated fermion KK tower are
very different from predictions made from an infinite fermion KK
tower. This apparent UV-sensitivity can actually be lifted by
considering the higher derivative operators described here (first
introduced in \cite{ATZ1}). When these are included, the predictions
made with a finite KK fermion tower become consistent with the
original predictions obtained with an infinite fermion tower.
A more esthetic problem with the brane Higgs scenario remains, since the definition of the
Higgs operators seems highly unnatural, if one understands a brane
Higgs field as a limit of a bulk Higgs field. All operators involving Higgs
fields will have to have a precise and definite dependence on $\beta$
(a large number), which seems quite contrived, specially in a framewrok in which no big
numerical hierarchies should arise from fundamental 5D
coefficients. In any case, with the ansatz outlined in the text
and reviewed in Appendix \ref{apen:II}, one can still work
consistently with a brane Higgs field as a limit case of a bulk Higgs field.

\acknowledgments
M.T. would like to thank Kaustubh Agashe for many discussions and
comments and specially Alex Azatov and Lijun Zhu for their invaluable
help and collaboration in the early stages of this work.
This work is supported in part by NSERC under grant number
SAP105354. 

\section{Appendix}
\label{sec:appendix}
\appendix

\section{Some Explicit Analytic Results }
\label{apen:I}

From equation (\ref{ym}) the shift defined as 
$\displaystyle \frac{\Delta}{m{v_4}}\equiv\frac{1}{{v_4}}-\frac{Y}{m}$, 
can be also derived from 
\bea
\frac{\Delta}{m{v_4}}={{v_4}\over Y^u_{qu}}
\sum_{i,j}{Y^u_{qU_j}Y^{u*}_{U_jQ_i}Y^u_{Q_iu}\over M_{Q_i}M_{U_j}}. 
\eea
Therefore simply replacing the second Yukawa coupling with the
 Yukawa coupling of the Higher derivative operator, $Y^R$ will give
\bea\label{YYY}
\frac{\Delta^R}{m{v_4}}={{v_4}\over Y^u_{qu}}
\sum_{i,j}{Y^u_{qU_j}Y^{Ru*}_{U_jQ_i}Y^u_{Q_iu}\over M_{Q_i}M_{U_j}}. 
\eea
Here, we present explicit analytic expressions for the $hgg$ production and also
the Yukawa coupling-mass shifts by performing the infinite sums over the KK modes. We  also include 
the result given in reference \cite{ATZ1} for the shift due to the usual 
Yukawa term, $Y_d\bar{Q}HU$, for completeness.\footnote{We have reproduced this 
result using the equation (\ref{YYY}), and our results match the one given
in the text of \cite{ATZ1}. Note however that there a few typos in the eq. A1 of their appendix.
}
To summarize, we have
\begin{eqnarray}
c_{hgg}&&=
\frac{2 m_d^2}{{v_4}} R'^2 \frac{2+c_u -c_q+\beta}{(1-2c_q)(1+2c_u)} 
\left [\frac{(1-\epsilon^{1-2c_q})(1-\epsilon^{1+2c_u})}{4+2\beta}
-\frac{1-\epsilon^{1-2c_q}}{5+2c_u+2\beta}
-\frac{1-\epsilon^{1+2c_u}}{5-2c_q+2\beta}
\right.
\nonumber\\
&&\left.
+\frac{\epsilon^{1-2c_q}}{4+2\beta}(1-\epsilon^{1+2c_u})
+\frac{\epsilon^{1+2c_u}}{4+2\beta}(1-\epsilon^{1-2c_q})
+\frac{\epsilon^{2-2c_q+2c_u}}{2-c_u+c_q+\beta}
-\frac{\epsilon^{-2c_q+1}}{3+c_u+c_q+\beta}
-\frac{\epsilon^{1+2c_u}}{3-c_u-c_q+\beta}
\right.\nonumber\\
&&\left.
+\frac{1}{4+c_u-c_q+\beta}
\right ]+{y_t^{RS}\over m_t}A_{1/2}(\tau_{t})+{\Delta_2^t\over m_t{v_4}}\, , 
\end{eqnarray}
for the $hgg$ production and
\begin{eqnarray}
\frac{\Delta^d_{1}}{m{v_4}}&&=
\frac{2 m_d^2}{{v_4}} R'^2 \frac{2+c_u -c_q+\beta}{(1-2c_q)(1+2c_u)} 
\left [\frac{(1-\epsilon^{1-2c_q})(1-\epsilon^{1+2c_u})}{6+c_u-c_q+3\beta}
-\frac{1-\epsilon^{1-2c_q}}{5+2c_u+2\beta}
-\frac{1-\epsilon^{1+2c_u}}{5-2c_q+2\beta}
\right.
\nonumber\\
&&\left.
+\frac{\epsilon^{1-2c_q}}{4+2\beta}(1-\epsilon^{1+2c_u})
+\frac{\epsilon^{1+2c_u}}{4+2\beta}(1-\epsilon^{1-2c_q})
+\frac{\epsilon^{2-2c_q+2c_u}}{2-c_u+c_q+\beta}
-\frac{\epsilon^{-2c_q+1}}{3+c_u+c_q+\beta}
-\frac{\epsilon^{1+2c_u}}{3-c_u-c_q+\beta}
\right.\nonumber\\
&&\left.
+\frac{1}{4+c_u-c_q+\beta}
\right ] \, ,
\end{eqnarray}
for the shifted Yukawa coupling. Also, there is a misalignment due to
the kinetic term \cite{ATZ1}, which 
as discussed in the text, is only important for the case of the third
generation quarks. We do not repeat 
that result here. For the higher derivative  term the shift is: 
\begin{eqnarray}\nonumber
\frac{\Delta^d_{R}}{m{v_4}}&&=
2\frac{Y'_R}{\Lambda^2}\frac{m_d^2}{{v_4}}\frac{2+c_u-c_q+\beta}{(1+2c_u)(1-2c_q)}
\left[
\frac{(4-c_q+\beta)(4+c_u+\beta)}{6+3\beta+c_u-c_q}(1-\epsilon^{1-2c_q})(1-\epsilon^{1+2c_u})
\right.\nonumber\\
&&\left.\
-\frac{(3-c_q)(4-c_q+\beta)}{5+2\beta-2c_q}(1-\epsilon^{1+2c_u})
-\frac{(3+c_u)(4+c_u+\beta)}{5+2\beta+2c_u}(1-\epsilon^{1-2c_q})
\right.\nonumber\\
&&\left.\
+\frac{(2+c_q)(4-c_q+\beta)}{4+2\beta}\epsilon^{1-2c_q}(1-\epsilon^{1+2c_u})
+\frac{(2-c_u)(4+c_u+\beta)}{4+2\beta}\epsilon^{1+2c_u}(1-\epsilon^{1-2c_q})
\right.\nonumber\\
&&\left.
+\frac{(2+c_q)(2-c_u)}{2+c_q-c_u+\beta}\epsilon^{2-2c_q+2c_u}
-\frac{(3-c_q)(2-c_u)}{3-c_u-c_q+\beta}\epsilon^{1+2c_u}
-\frac{(2+c_q)(3+c_u)}{3+c_u+c_q+\beta}\epsilon^{1-2c_q}
+\frac{(3-c_q)(3+c_u)}{4+c_u-c_q+\beta}
\right].\nonumber
\end{eqnarray}

\section{From Bulk to Brane}
\label{apen:II}
%
We summarize the matching prescription for operators containing Higgs field for the case where the Higgs boson is localized on the brane. As explained in Section \ref{sec:setup}, these prescriptions insures that the 5D bulk Higgs scenario transitions smoothly to a brane-localized Higgs case. 
The brane prescription for the Higgs associates a delta function to
the Higgs normalization integral 

$$\int_R^{R' }(\frac{R}{z})^3 dz [h_\beta(z)]^2=1\,.$$
As the $HH$, rather than $H$ field,  is associated with a $\delta$
function, one must include a $\beta$ dependence to the bulk Higgs
fields to be able to match operators, in the limit $\beta \to \infty$
to the brane ones. The conversion is:
\begin{eqnarray}
&&H \to \sqrt{\beta}\, , \nonumber\\
&&HH \to HH \, , \nonumber \\
&&HHH \to \frac{1}{\sqrt{ \beta}}HHH \, , \nonumber
\end{eqnarray}
 for matching brane to bulk in the appropriate limit.

So for the shift, we have contributions from $Y_2$ and $Y_R$. As we
are dealing with an effective theory, we look at the effect of summing
over a finite number of modes, let's say 3 to 5.  

For the case of brane Higgs, the contributions for a finite number of
modes for $Y_2$ give exactly $0$ (because of boundary values on the
brane). This confirms the work of \cite{Bouchart:2009vq}. However, we
must add higher order operators $Y_R$, which give a significant result
(converging to a constant for $\beta \to 1000$ and anything
beyond). The result obtained by summing over a finite number of modes
in the brane on the $Y_R$ contribution must be compared with the
result in the paper by \cite{ATZ1} for the infinite sum of $Y_2$ on
the brane. 

For bulk Higgs, the shift contribution from a finite number of modes
on the  $Y_2$ contribution is no longer $0$. However, adding to this
the $Y_R$ contribution, we notice that the $Y_R$ contribution for bulk
Higgs is much smaller (two orders of magnitude) than the corresponding
one in the brane. This is a clear indication that higher order
corrections are much more important for the brane Higgs case than for
the bulk.


\end{document}